\documentclass[12pt]{article}
\usepackage{graphicx}
\usepackage{indentfirst}
\usepackage{multicol}
\usepackage{amsmath}
\usepackage{pstricks}
\newcommand{\ad}{^\dagger }
\newcommand{\adp}{^{\phantom{\dagger}}}

\newcommand{\becs}{\begin{cases}}
\newcommand{\bem}{\begin{matrix}}
\newcommand{\blp}{\bigl(} 
\newcommand{\Blp}{\Bigl(} 
\newcommand{\Brp}{\Bigr)}
\newcommand{\brp}{\bigr)}

\newcommand{\encs}{\end{cases}}
\newcommand{\enm}{\end{matrix}}

\newcommand{\hquad}{\mspace{8 mu}} 

\newcommand{\lgl}{\langle } 
\newcommand{\lp}{\left(} 

\newcommand{\ot}{\otimes }

\newcommand{\ra}{\rightarrow }

\newcommand{\rgl}{\rangle }

\newcommand{\rp}{\right)}

\newcommand{\tm}{\times }
\newcommand{\Tr}{{\rm Tr}}

\newcommand{\vb}{\,|\,}

\newcommand{\HS}{{\cal H}}

\newcommand{\al}{\alpha }
\newcommand{\bt}{\beta }
\newcommand{\gm}{\gamma }

\newcommand{\dl}{\delta }

\newcommand{\ep}{\epsilon}

\newcommand{\kp}{\kappa }
\newcommand{\lm}{\lambda }

\begin{document}

\title{Deterministic and Unambiguous Dense Coding}

\author{ 
Shengjun Wu$^{1,2}$, Scott M. Cohen$^{1,3}$, Yuqing Sun$^1$, 
Robert B. Griffiths$^1$\\
$^1$Physics Department,
Carnegie-Mellon University,\\
Pittsburgh, PA 15213, USA\\
$^2$Hefei National Laboratory for Physical Science at the Microscale,\\
University of Science and Technology of China,\\
Hefei, Anhui 230026, P. R. China\\
$^3$Physics Department,
Duquesne University,
Pittsburgh, PA 15282, USA}

\date{Version of 14 February 2006}

\maketitle  

\begin{abstract}

Optimal dense coding using a partially-entangled pure state of Schmidt rank
$\bar D$ and a noiseless quantum channel of dimension $D$ is studied both in
the deterministic case where at most $L_d$ messages can be transmitted with
perfect fidelity, and in the unambiguous case where when the protocol succeeds
(probability $\tau_x$) Bob knows for sure that Alice sent message $x$, and when
it fails (probability $1-\tau_x$) he knows it has failed.  Alice is allowed
any single-shot (one use) encoding procedure, and Bob any single-shot
measurement.

For $\bar D\leq D$ a bound is obtained for $L_d$ in terms of the largest
Schmidt coefficient of the entangled state, and is compared with published
results by Mozes et al.  For $\bar D > D$ it is shown that $L_d$ is strictly
less than $D^2$ unless $\bar D$ is an integer multiple of $D$, in which case
uniform (maximal) entanglement is not needed to achieve the optimal protocol.

The unambiguous case is studied for $\bar D \leq D$, assuming $\tau_x>0$ for a
set of $\bar D D$ messages, and a bound is obtained for the average
$\lgl1/\tau\rgl$.  A bound on the average $\lgl\tau\rgl$ requires
an additional assumption of encoding by isometries (unitaries when $\bar D=D$)
that are orthogonal for different messages.  Both bounds are saturated when
$\tau_x$ is a constant independent of $x$, by a protocol based on one-shot
entanglement concentration. For $\bar D > D$ it is shown that (at least) $D^2$
messages can be sent unambiguously. 

Whether unitary (isometric) encoding suffices for optimal protocols remains a
major unanswered question, both for our work and for previous studies of dense
coding using partially-entangled states, including noisy (mixed) states.

\end{abstract}

\section{Introduction}
\label{sct1}

Dense coding is an intriguing nonclassical effect made possible by entangled
quantum states: combining entanglement with a quantum channel allows more
information to be transmitted than is possible using these resources
separately.  The original protocol of Bennett and Wiesner \cite{BnWs92} can be
summarized, in slightly altered notation, as follows.  Alice and Bob share two
$D$-dimensional particles, meaning that each is described in quantum terms
using a $D$-dimensional Hilbert space, which are initially in a fully-entangled
state.  Alice carries out one of $D^2$ mutually-orthogonal unitary
encoding operations on her particle and sends it to Bob through a
perfect $D$-dimensional quantum channel. Bob measures the quantum state of the
two-particle in a fully-entangled orthonormal basis in order to learn with
certainty which of the $D^2$ operations Alice carried out.  This protocol
can transmit $D^2$ ``classical'' messages with perfect fidelity, corresponding
to a classical channel of capacity $2\log D$.  For a careful and mathematically
precise discussion of what we shall hereafter refer to as the \emph{standard
protocol}, see \cite{Wrnr01}.

If the two particles are in a partially, as opposed to fully, entangled
state, can dense coding still be carried out, and if so, how many messages can
be transmitted?  How must the standard protocol be modified in order to
accomplish this?  Even if $D^2$ messages cannot be sent with certainty, are
there probabilistic protocols which allow significantly more information to be
transmitted than the $\log D$ capacity of the quantum channel by itself?
These questions have led to a significant body of research.
The present paper addresses them in two particular cases.

The first is \emph{deterministic dense coding}, where the aim is to
send $L$ distinct messages with perfect fidelity, and a significant problem is
to determine the maximum value $L_d$ of $L$ for a given partially-entangled
state.  The answer is known for a \emph{uniformly entangled state}, our term
for a state in which all the nonzero Schmidt coefficients are identical, when
the Schmidt rank $\bar D$ is \emph{less} than the channel dimension $D$: an
appropriate modification of the standard unitary encoding protocol allows the
transmission of $L_d=\bar DD$ messages.  For other situations few exact results
are available, though a number of interesting numerical and analytical results
have recently been published by Mozes et al.\ \cite{MzOR05}.
Our contribution to this topic consists in part in raising the question, which
we are unable to answer, as to whether unitary (or isometric, see
Sec.~\ref{sct2}) encoding is sufficient to achieve the maximum value $L_d$.  In
the case of $\bar D$ less than $D$ we derive a rigorous inequality for $L_d$
which holds for a general encoding protocol, unitary or not, and compare it
with some of the results in \cite{MzOR05}.  We also explore, in a preliminary
way, the situation when $\bar D$ (the Schmidt rank) is larger than $D$ (the
dimension of the quantum channel), for which unitary encoding is impossible,
and show that $L_d$ is strictly less than $D^2$ unless $\bar D$ is an integer
multiple of $D$.

The second case we consider is \emph{unambiguous dense coding}: when Alice
encodes message $x$, Bob's measurement will with probability $\tau_x$ tell him
precisely which message Alice sent, and with probability $1-\tau_x$ that the
protocol has failed.  A significant problem is to determine the maximum average
probability of success $P_s=\lgl\tau\rgl$ for some set of $L$ messages.  This
will depend on the choice of $L$, with $P_s$ decreasing, for a given entangled
state, as $L$ increases.  We consider the case $L=\bar D D$, the maximum number
of messages that can be sent in unambiguous fashion for $\bar D \leq D$, and
derive a bound for the the average \emph{inverse} probability of success
$\lgl1/\tau\rgl$, assuming $\tau_x>0$ for every message.  We also obtain a
bound for $\lgl\tau\rgl$ when encoding is carried out using \emph{orthogonal}
isometries (orthogonal unitaries in the case $\bar D=D$).  Both bounds are
saturated in the special case in which $\tau_x$ is independent of $x$ by a
protocol which employs unambiguous entanglement concentration.  There are many
other cases one might wish to consider, and for these our bounds are less
useful.  In particular, the situation when the number $L$ of messages with
$\tau_x>0$ is less than $\bar D D$ is hard to analyze, because one cannot be
sure that unitary (or isometric) encoding is the optimal strategy.  Indeed, the
issue of determining when unitary encoding is optimal remains a major
unanswered question in studies of dense coding using partially entangled
states. In our opinion it deserves a lot more attention than it has hitherto
received.  We can say little about unambiguous protocols for $\bar D> D$ aside
from showing that it is always possible to send $D^2$ messages with positive
probability.  We suspect this is the maximum possible number, but we have no
proof, nor can we identify an optimal protocol.

The remainder of this paper is organized in the following
order. Section~\ref{sct2} introduces our notation for entangled states,
encoding operations, and measurement POVMs appropriate for deterministic and
unambiguous protocols.  Next we derive, in Sec.~\ref{sct3}, some very general
information-theoretic bounds, which are later compared with our inequalities
based on Schmidt coefficients of the entangled state.  Deterministic dense
coding is the topic of Sec.~\ref{sct4}: first a review in \ref{sct4a} of the
case of uniformly-entangled states for $\bar D\leq D$, next in \ref{sct4b} a
rigorous inequality for $L_d$, followed in \ref{sct4c} by a discussion of what
happens for $\bar D>D$, and in \ref{sct4d} by a remark on protocols that
achieve $L_d$.  Our discussion of unambiguous dense coding for $\bar D\leq D$
begins in Sec.~\ref{sct5} with the derivation of the inequality for $\lgl
1/\tau\rgl$, and continues in Sec.~\ref{sct6}, where additional results are
obtained assuming encoding using orthogonal isometries. We derive a bound on
$\lgl\tau\rgl$ in \ref{sct6a}, and show in \ref{sct6b} that such encoding is
optimal when $\tau_x$ is independent of $x$ for a set of $\bar DD$ messages.
However, the bound on $\lgl\tau\rgl$ given in \ref{sct6c} is unlikely to hold
if one allows nonorthogonal isometries.  Unambiguous dense coding in a
situation with $\bar D > D$ is the subject of Sec.~\ref{sct7}.

Section~\ref{sct8} contains material relating our work to previous research:
uniform entanglement for $\bar D <D$ in \ref{sct8a}; deterministic dense coding
with reference to \cite{MzOR05} in \ref{sct8b}; previous work on unambiguous
dense coding in \ref{sct8c}; the connection of unambiguous dense coding with
unambiguous discrimination in \ref{sct8d}; and finally, the issue of unitary
encoding when carrying out dense coding using noisy (mixed) entangled states,
in \ref{sct8e}.  The concluding Sec.~\ref{sct9} contains a summary of our
results followed by a discussion of open questions.  Two appendices contain
technical results.

	\section{General Framework}
\label{sct2}

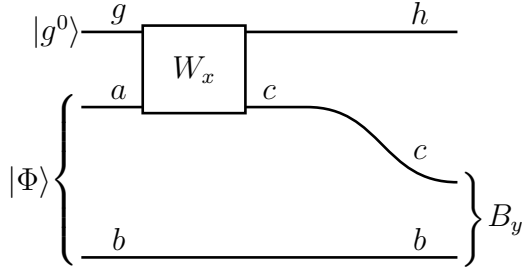
\begin{figure}[h]
$$
\begin{pspicture}(-0.7,-0.5)(6,3.5) 
\newpsobject{showgrid}{psgrid}{subgriddiv=1,griddots=10,gridlabels=6pt}
\def\lwd{0.035} 
\psset{
labelsep=2.0, arrowsize=0.150 1,linewidth=\lwd} 
\def\vvv#1{\vrule height #1 cm depth #1 cm width 0pt}
\def\rectc(#1,#2){%
\psframe[fillcolor=white,fillstyle=solid](-#1,-#2)(#1,#2)}
\def\squ{\rectc(0.35,0.35)}
\psline(0.0,0.0)(5.0,0.0)
\psline(0.0,3.0)(5.0,3.0)
\psline(0.0,2.0)(3.0,2.0)
\psbezier(3.0,2.0)(4.0,2.0)(4.0,1.0)(5.0,1.0)
\rput(1.5,2.5){\rectc(0.7,0.6)}
\rput(1.5,2.5){$W_x$}
\rput[r](0.0,3.0){$|g^0\rgl$}
\rput[r](0,1){$\left\{\vvv{1.1}\right.$}
\rput[r](-0.4,1){$|\Phi\rgl$}
\rput[l](5.0,0.5){$\left.\vvv{0.5}\right\}$}
\rput[l](5.4,0.5){$B_y$}
\rput[b](0.5,3.1){$g$}
\rput[b](4.5,3.1){$h$}
\rput[b](0.5,2.1){$a$}
\rput[b](2.5,2.1){$c$}
\rput[b](4.5,1.3){$c$}
\rput[b](0.5,0.1){$b$}
\rput[b](4.5,0.1){$b$}
\end{pspicture}
$$
\caption{Encoding and measurement for a dense coding protocol.
 }
\label{fgr1}
\end{figure}

The general setup we shall be studying is shown schematically in
Fig.~\ref{fgr1}. Alice and Bob share a normalized entangled state
\begin{equation}
  |\Phi\rgl = \sum_{j=1}^{\bar D} \lm_j |a^j\rgl\ot|b^j\rgl
\label{eqn1}
\end{equation}
of Schmidt rank $\bar D$ on the tensor product $\HS_a\ot\HS_b$ of two Hilbert
spaces of dimension $d_a$ and $d_b$, with orthonormal bases $\{|a^j\rgl\}$ and
$\{|b^j\rgl\}$. We assume the Schmidt coefficients are ordered from largest to
smallest,
\begin{equation}
  \lm_1\geq \lm_2\geq \cdots \lm_{\bar D} > 0,
\label{eqn2}
\end{equation}
and $\lm_j=0$ for $j>\bar D$. One can visualize the situation by assuming that
$\HS_a$ and $\HS_b$ refer to two particles, one in Alice's and one in Bob's
possession.  Our analysis is simplified through assuming that
\begin{equation}
  d_a=d_b=\bar D,
\label{eqn3}
\end{equation}
which is to say $|\Phi\rgl$ is of full Schmidt rank on $\HS_a\ot\HS_b$, but
nothing essential would change if $d_a$ or $d_b$ had values larger than $\bar
D$.  In addition, Alice can signal Bob through a perfect $D$-dimensional
quantum channel, where $D$ does \emph{not} have to be the same as $\bar D$.

Alice wishes to transmit one of $L$ messages labeled $x=1, 2, \ldots$ to Bob,
and to do so she \emph{encodes} her message by carrying out a unitary map $W_x$
from $\HS_g\ot \HS_a$ to $\HS_h\ot\HS_c$, where $\HS_g$ refers to an ancillary
particle in a pure state $|g^0\rgl$, $\HS_h$ to the final ancillary
particle, and $\HS_c$ is a Hilbert space of dimension $d_c=D$, thought of as a
particle which is then sent through the noiseless channel to Bob.  Since $W_x$
is unitary, $d_g d_a=d_h d_c$, but $d_a=\bar D$ need not be the same as
$d_c=D$.  By introducing an orthonormal basis $\{|h^l\rgl\}$ for $\HS_h$, we
can express the action of $W_x$ in the form
\begin{equation}
  W_x\Blp|g^0\rgl\ot |a\rgl\Brp = \sum_l |h^l\rgl\ot\Blp A_{xl}|a\rgl\Brp,
\label{eqn4}
\end{equation}
where the $A_{xl}$, which are known as Kraus operators, map $\HS_a$ to $\HS_c$
and satisfy the normalization condition
\begin{equation}
  \sum_l A_{xl}\ad A\adp_{xl} = I_a.
\label{eqn5}
\end{equation}

In the special case in which there is only a single term $l=1$ in this sum, we
will omit the subscript $l$ and refer to the map $A_x$ of $\HS_a$ to $\HS_c$ as
an \emph{isometry}, since $A_x\ad A_x=I_a$ means that $A_x$ preserves norms.
An isometry is only possible when $d_a=\bar D\leq D=d_c$, and if $\bar D=D$ the
isometry is a unitary operator.  In this sense \emph{isometric encoding}
represents a natural generalization of unitary encoding in the standard
protocol.  To be sure, when $\bar D$ is less than $D$ one can always
suppose that $d_a=D$ and that $|\Phi\rgl$ is supported on a subspace of
$\HS_a$, so that the particle $c$ sent through the channel is identical with
the particle $a$ initially in an entangled state.  However, we find it more
convenient to carry out the analysis assuming $d_a=\bar D$. If $\bar D$ is
larger than $D$, one must assume different dimensions for $\HS_a$ and $\HS_c$,
and isometric encoding is not possible. 

Bob's task is to extract information by carrying out a POVM using a collection
$\{B_y\}$ of positive operators on $\HS_c\ot\HS_b$, as indicated schematically
in Fig.~\ref{fgr1}.  For studying unambiguous dense coding it is convenient to
assume that the label $y$ can take on values $0, 1, 2, \ldots,$ with the
significance that if the outcome is $y=x>0$, Bob knows for sure that Alice sent
message $x$, while $y=0$ is the ``failure'' or ``garbage'' outcome: he does not
know which message was sent.  As is well known, Bob's POVM can always be
thought of as a projective measurement on $\HS_c\ot\HS_b$ along with an
ancillary system prepared in a pure state, and the reader may wonder why we
have not included this ancillary system as part of Fig.~\ref{fgr1}.  The answer
is that it is not needed for our analysis, whereas the details of Alice's
encoding procedure play a more significant role in our discussion.

The framework outlined above for encoding and decoding is also appropriate,
given some obvious modifications, for the case in which Alice and Bob share a
noisy entangled state, represented by a density operator or ensemble, or use a
noisy channel.  But our entire discussion is limited to ``one
shot'' dense coding: Alice does not entangle her input over many uses of the
apparatus, nor does Bob save the outcomes of multiple transmissions in order to
perform a coherent measurement.

An unambiguous dense coding protocol is thus one in which the $\{W_x\}$---or
equivalently the $\{A_{xl}\}$---and the $\{B_y\}$ have been chosen so that
\begin{equation}
  \Pr(y\vb x) = \tau_x\dl_{yx} +(1-\tau_x)\dl_{y0},
\label{eqn6}
\end{equation}
where $\tau_x$ is the probability that if Alice chooses to send message $x$ it
will be correctly transmitted: Bob's apparatus will show $y=x$ rather than
$y=0$, the latter being an indication that the protocol has failed.  If
$\eta_x$ is the a priori probability for Alice choosing message $x$, the joint
probability distribution will be
\begin{equation}
  \Pr(x,y) = \eta_x\left[ \tau_x\delta _{yx} + (1-\tau_x)\delta _{y0}\right].
\label{eqn7}
\end{equation}
This means the average probabilities $P_s$ of success and $P_f$ of failure in
sending a message are
\begin{equation}
  P_s = \sum_x \eta_x\tau_x,\quad P_f = 1-P_s.
\label{eqn8}
\end{equation}

The \emph{deterministic} case is one in which $\tau_x=1$ for $1\leq x\leq L$,
where $L$ is the number of messages under consideration, and an \emph{optimal}
deterministic protocol is one giving rise to the maximum number $L_d$ of
messages having $\tau_x=1$, assuming $|\Phi\rgl$ and $D$ are held fixed.  An
optimal unambiguous protocol is, roughly speaking, one that yields the maximum
value of $P_s$ but this will depend on the a priori probabilities $\{\eta_x\}$.
We shall only consider the case $\eta_x = 1/L$.

	\section{Information Theory Bound}
\label{sct3}

Before discussing specific protocols in the following sections, it is
convenient to derive some simple but quite general information-theoretic bounds
on the probability of successfully transmitting a message from Alice to Bob.
The first is based on the result in Sec.~VII of \cite{Hsao96}, that the
classical capacity $C$ of a dense coding ``channel'' (entangled state plus
quantum channel) of the sort we are considering is given by
\begin{equation}
  \bar D\leq D:\; C = \log D +H_E,\quad H_E=-\sum_j \lambda _j^2\log\lambda
_j^2;
\label{eqn9}
\end{equation}
$H_E$ is the entanglement of $|\Phi\rgl$. The condition $\bar D\leq D$ is
implicit in the derivation in \cite{Hsao96}.  For $\bar D > D$ we do not know
of a comparable expression, but studies of entanglement assisted capacity
\cite{BSST02} yield an upper bound
\begin{equation}
  \bar D > D:\; C\leq 2 \log D.
\label{eqn10}
\end{equation}
(This also holds for $\bar D\leq D$, but then it is obvious from
\eqref{eqn9}, since $H_E$ cannot exceed $\log D$.)

A consequence of \eqref{eqn7} is the conditional probability
\begin{equation}
  \Pr(x\,|\, y)=\begin{cases} \delta _{yx} &\text{ for $y>0$,}\\
  \displaystyle \frac{\eta_x(1-\tau_x)}{1-P_s} &\text{ for $y=0$,}
\end{cases}
  \label{eqn11}
\end{equation}
from which the Shannon mutual information
\begin{equation}
  I(X\,\hbox{:}\, Y) = H(X) - H(X\,|\, Y),
\label{eqn12}
\end{equation}
can be calculated using
\begin{equation}
\begin{split}
  H(X) &= -\sum_x\eta_x\log\eta_x,\\ H(X\,|\, Y) &= (1-P_s) H(X\,|\, y=0).
\end{split}
\label{eqn13}
\end{equation}
Because of \eqref{eqn11}, $H(X\,|\, y)=0$ for all $y>0$.

If we restrict ourselves to the situation in which all $L$ messages have the
same \textit{a priori} probability, $\eta_x=1/L$, and use the upper bound
$H(X\,|\, y=0)\leq \log L$ in \eqref{eqn13}, the fact that $I(X:Y)$ cannot
exceed $C$ in
\eqref{eqn9} leads to the inequality
\begin{equation}
  P_s\log L \leq \log D + H_E
\label{eqn14}
\end{equation}
 with
\begin{equation}
  P_s = (1/L)\sum_{x\geq 1} \tau_x
\label{eqn15}
\end{equation}
the average unweighted probability of successfully transmitting a message.

In certain cases this bound might be improved by choosing a nonuniform set of a
priori probabilities $\{\eta_x\}$, or using a better upper bound than $\log L$
for $H(X\,|\, y=0)$.  However, because of their generality, one cannot expect
bounds of this sort to be very tight, and in the following sections we obtain
for restricted types of protocols improved bounds which are not based on
Shannon mutual information.

	\section{Deterministic Dense Coding}
\label{sct4}

	\subsection{Uniformly entangled state with $\bar D \leq D$}
\label{sct4a}

We use the term \emph{uniformly entangled} for the state $|\Phi\rgl$ in
\eqref{eqn1} when all the (nonzero) $\lm_j$ are equal to each other.  When
$\bar D=d_a=d_b$ this coincides with the terms ``fully'' or ``maximally
entangled,'' but neither term seems appropriate when $\bar D$ is smaller.  One
of the simplest and most straightforward extensions of the standard dense
coding scheme is to a uniformly entangled state $\bar D < D$, which can be used
to send exactly $\bar DD$ messages deterministically when encoded using
orthogonal isometries.

Let $\{A_x\}$ be a collection of isometries from $\HS_a$ to $\HS_c$, see the
discussion following \eqref{eqn5}, which are \emph{orthogonal} in the sense
that
\begin{equation}
  \Tr_a(A\ad_x A\adp_y) = \bar D\dl_{xy} = \Tr_c(A\adp_yA\ad_x),
\label{eqn16}
\end{equation} 
where one can use either equation as a definition.  If we use the orthonormal
basis $\{|a^j\rgl\}$ in \eqref{eqn1} to write each $A_x$ in the form
\begin{equation}
   A_x = \sum_{j=1}^{\bar D} |\gm^j_x\rgl\lgl a^j|,
\label{eqn17}
\end{equation}
where the expansion coefficients $|\gm^j_x\rgl$ are elements of $\HS_c$, the
orthogonality condition \eqref{eqn16} becomes
\begin{equation}
  \sum_{j=1}^{\bar D} \lgl\gm^j_x|\gm^j_y\rgl = \bar D\dl_{xy}.
\label{eqn18}
\end{equation}
It follows from this that if $|\Phi\rgl$ is uniformly entangled, the kets
\begin{equation}
  |\Phi_x\rgl = A_x|\Phi\rgl = \sum_j \Blp1/\sqrt{\bar
D}\,\Brp\,|\gm^j_x\rgl\ot|b^j\rgl
\label{eqn19}
\end{equation}
on $\HS_c\ot\HS_b$ form an orthonormal collection. Given a uniformly entangled
state and a collection of $\bar DD$ orthogonal isometries, there is a
straightforward deterministic dense coding protocol for $L=\bar DD$ messages:
Alice uses $A_x$ to encode message $x$, and Bob measures using the orthonormal
basis $\{|\Phi_x\rgl\}$.  A proof of the intuitively obvious result that such a
protocol is optimal can be based on \eqref{eqn14} with $H_E=\log\bar D$, or on
\eqref{eqn26} below. Of course, when $\bar D=D$ the isometries are unitaries,
and we are back to the standard protocol.

Here is one way to construct $\bar DD$ orthogonal isometries.  Let operators
$Q:\HS_a\ra\HS_c$, $R:\HS_a\ra\HS_a$, and $S:\HS_c\ra\HS_c$ be defined by
\begin{equation}
  R|a^j\rgl = e^{2\pi i(j-1)/\bar D}|a^j\rgl,\quad S|c^k\rgl =
|c^{k\oplus1}\rgl, \quad Q = \sum_{j=1}^{\bar D} |c^j\rgl\lgl a^j|
\label{eqn20}
\end{equation}
in terms of orthonormal bases $\{|a^j\rgl\}$, $1\leq j\leq \bar D$, of $\HS_a$
and $\{|c^k\rgl\}$, $1\leq k\leq D$, of $\HS_c$; $\oplus$ means addition modulo
$D$. Since $R$ and $S$ are unitary, each of the $\bar DD$ operators
\begin{equation}
  A_{\al\bt} = S^\bt Q R^\al,\quad 0\leq \al < \bar D,\hquad 0\leq \bt < D,
\label{eqn21}
\end{equation}
is an isometry from $\HS_a$ to $\HS_c$.  Here each distinct double subscript
${\al\bt}$ corresponds to a different value of $x$, and the counterpart of
\eqref{eqn16} is
\begin{equation}
  \Tr_a(A_{\al\bt}\ad A_{\al'\bt'}\adp) = \bar D\dl_{\al\al'}\dl_{\bt\bt'}.
\label{eqn22}
\end{equation}

	\subsection{General bound $L_d\leq D/\lm_1^2$}
\label{sct4b}

Whatever operation Alice carries out to encode message $x$ will result in a
density operator $\rho_x$ describing the combined $\HS_c\ot\HS_b$ system which
Bob will measure, see Fig.~\ref{fgr1}, and since whatever Alice does has no
effect on Bob's particle, its reduced density operator is
\begin{equation}
  \Tr_c(\rho_x) =\Tr_a\blp |\Phi\rgl\lgl\Phi|\brp = \sum_j \lm_j^2 |b^j\rgl\lgl
b^j|,
\label{eqn23}
\end{equation}
independent of $x$.  Two density operators $\rho_x$ and $\rho_y$ corresponding
to distinct messages $x$ and $y$ can only be distinguished with certainty
\cite{FnDY04} if $\rho_x\rho_y=0$, which is to say their supports are
orthogonal: $P_x P_y=0$, where $P_x$ is the projector onto the support of
$\rho_x$.  Consequently, since $\rho_x\leq P_x$ in the sense that $P_x-\rho_x$
is a positive operator, for a deterministic protocol it must be the case that
\begin{equation}
  \sum_{x=1}^L \rho_x\leq \sum_x P_x \leq I_c\ot I_b.
\label{eqn24}
\end{equation}
Upon tracing this inequality over $\HS_c$ and using \eqref{eqn23}, one obtains
\begin{equation}
  L\sum_j \lm_j^2 |b^j\rgl\lgl b^j| \leq DI_b = D\sum_j |b^j\rgl\lgl b^j|,
\label{eqn25}
\end{equation}
so that $L\lm_j^2\leq D$ for every $j$.  Since $\lm_1$ is the largest Schmidt
coefficient of $|\Phi\rgl$, this implies that the maximum value of $L$
satisfies
\begin{equation}
   L_d \leq D/\lm_1^2.
\label{eqn26}
\end{equation}

As $\lm_1^2$ cannot be smaller than $1/\bar D$, this inequality implies that
$L_d$ cannot exceed $\bar DD$, a bound which is achievable for $\bar D \leq D$
using a uniformly entangled state, as shown in part A, but not for $\bar D >
D$, see part C below. If $|\Phi\rgl$ is a product state, $\lm_1=1$ and the
rather trivial bound $L_d\leq D$ is achieved by sending one of $D$ orthogonal
states through the quantum channel. In other situations the bound \eqref{eqn26}
is less trivial; see, in particular, the discussion in Sec.~\ref{sct8b}.

Taking the logarithm of \eqref{eqn26}, one has
\begin{equation}
  \log L_d\leq \log D + \log(1/\lm_1^2)\leq \log D + H_E,
\label{eqn27}
\end{equation}
where the second inequality follows from the definition in \eqref{eqn9}, given
that $\lm_1^2\geq\lm_j^2$ for all $j$ and $\sum\lm_j^2=1$. This shows that
\eqref{eqn26} is a tighter bound than the information-theoretic \eqref{eqn14}
with $L=L_d$ and $P_s=1$.

	\subsection{Protocols for $\bar D > D$}
\label{sct4c}

The case $\bar D > D$ stands in marked contrast with that for $\bar D\leq D$
discussed in part A above.  To begin with, it is impossible to send $\bar DD$
messages in a deterministic fashion, because that exceeds the bound $D^2$
implied by \eqref{eqn10}.  But even sending $D^2$ messages is not possible
unless $\bar D$ is an integer multiple of $D$; otherwise, as we shall show,
$L_d$ is strictly less that $D^2$.  Furthermore, when $\bar D$ is an integer
multiple of $D$, a uniformly entangled state is \emph{not} needed to achieve
the optimal protocol, though there is still a nontrivial constraint on the
Schmidt coefficients.

Let us first discuss the case $D=2$, $\bar D=4$, assuming $|\Phi\rgl$ is a
uniformly entangled state.  Without loss of generality one can think of this
(up to some local unitaries) as Alice and Bob sharing two fully-entangled qubit
pairs.  An optimal dense coding protocol consists in throwing away one pair,
and carrying out standard dense coding with the other, in order to send one of
$D^2=4$ messages.  But of course the pair that was thrown away need not have
been fully entangled, so it is at least sufficient that the Schmidt
coefficients be identical in pairs: $\lm_1=\lm_2$, and $\lm_3=\lm_4$.  Indeed,
the pair that was thrown away could have been in a mixed state.  The general
case in which $\bar D$ is an integer multiple of $D$ can be discussed in
exactly the same way whenever $|\Phi\rgl$ can be thought of as the tensor
product of one fully-entangled $D\tm D$ pair with something else: by discarding
the latter, which could have been in any state whatsoever, and using the former
for standard dense coding one achieves an optimal protocol.  Describing all of
this in terms of the Kraus operators and the POVM of Sec.~\ref{sct2} is an
exercise we leave to the reader.

Next assume that $\bar D$ is \emph{not} a multiple of $D$. Alice must encode a
particular message $x$ using a collection of Kraus operators $\{A_{xl}\}$
satisfying \eqref{eqn5}. Since $A_{xl}$ maps a $\bar D$-dimensional space to
one of dimension $D<\bar D$, its rank, which is the same as the rank of
$A_{xl}\ad A^{}_{xl}$ (p.~13 of \cite{HrJh85}), is at most $D$.  But the
identity operator $I_a$ in \eqref{eqn5} is of rank $\bar D> D$. Therefore
encoding cannot be achieved using a single Kraus operator, but requires a Kraus
rank $\kp$ (number of independent Kraus operators) bounded below by
\begin{equation}
  \kp \geq \xi := \lceil \bar D/D\rceil,
\label{eqn28}
\end{equation}
where $ \lceil \al\rceil$ is the smallest integer not less than $\al$.  This in
turn has the consequence, as shown in App.~\ref{sctpa}, that Alice's encoding
results in a state which when traced down to the $\HS_c\ot\HS_b$ space
available to Bob corresponds to a density operator $\rho_x$ of rank greater
than or equal to $\kp$, whose support is therefore a subspace of dimension at
least $\kp$.  As noted above in B, two density operators $\rho_x$ and $\rho_y$
can be distinguished with certainty if and only if their supports are
orthogonal, and this means that the number of messages that can be sent
deterministically is bounded above by
\begin{equation}
  L_d\leq \mu:=\lfloor \bar DD/\xi\rfloor,
\label{eqn29}
\end{equation}
where $\lfloor \al\rfloor$ denotes the largest integer not greater than $\al$.

When $\bar D$ is an integer multiple of $D$, $\mu=D^2$, and, as shown earlier,
one can achieve this value for $L_d$ by using an appropriate $|\Phi\rgl$.
However, if $D$ does not divide $\bar D$, $\mu$ will lie somewhere in the range
\begin{equation}
  D(D+1)/2\leq \mu < D^2,
\label{eqn30}
\end{equation}
so $L_d$ is less than $D^2$, a result which is tighter than the
information-theoretic bound \eqref{eqn10}.  But we do not know whether
$L_d=\mu$ can actually be achieved, even in the simplest case in which $D=2$,
$\bar D=3$, for which $\xi=2$ and $\mu=3$.  That is, we have been unable to
design a deterministic protocol for transmitting 3 messages, or to show that it
is impossible.  If for this case only 2 messages can be sent deterministically,
the entangled state is of no use and might as well be thrown away.

The argument that produces the bound in \eqref{eqn29} does not require that
$|\Phi\rgl$ be uniformly entangled, but only that it have Schmidt rank $\bar
D$.  There are, of course, cases in which Nielsen's majorization condition
\cite{Nlsn99} will permit such a state to be replaced with probability 1 by a
uniformly entangled state of rank $D$, which would allow $L_d=D^2$ messages to
be sent deterministically using the standard protocol. However, the replacement
requires both local operations \emph{and} classical communication, which in the
dense coding context means a classical side channel.  That lies outside the
scope of the present paper, though it belongs to a class of problems worthy of
further exploration.

	\subsection{No extension of optimal deterministic protocol}
\label{sct4d}

Suppose a deterministic protocol is \emph{optimal} in the sense that
\begin{equation}
  \tau_x=1 \text{ for } 1\leq x\leq L_d,
\label{eqn31}
\end{equation}
with $L_d$ the maximum possible number of deterministic messages for a given
$|\Phi\rgl$, $\bar D$ and $D$.  It is then not possible to find an unambiguous
protocol in which the same number of messages can be sent deterministically,
and \emph{ in addition} one or more messages can be sent unambiguously with
probabilities of success less than 1.  In other words,
\eqref{eqn31} implies that $\tau_x=0$ for $x>L_d$. 

The argument is straightforward.  The density operators (possibly pure states)
$\rho_x$ created by Alice on $\HS_c\ot\HS_b$ must be mutually orthogonal, see
the arguments in part C above, for $1\leq x\leq L_d$. Were it possible for her
to create yet another $\rho'$ for an additional message $x=L_d+1$, it would
have to be orthogonal to all the $\rho_x$ just mentioned, as otherwise there
would be at least one message in the set $1\leq x\leq L_d$ which Bob could not
definitely distinguish from message $L_d+1$.  But if $\rho'$ were orthogonal in
this way, the additional message could also be sent deterministically, contrary
to the assumption that $L_d$ is the maximum number possible.

	\section{Saturated Unambiguous Dense Coding}
\label{sct5}

Whereas unambiguous dense coding is a complicated problem if one allows the
most general encoding and decoding protocols, the situation is
considerably simpler for $\bar D\leq D$ if one supposes that precisely $\bar
DD$ messages can be unambiguously transmitted, each with a
\emph{positive} (conditional) probability $\tau_x$.

To begin with, it is impossible to send \emph{more} than $\bar DD$ messages,
because Bob is carrying out measurements on a $\bar DD$-dimensional Hilbert
space, and very general arguments \cite{Chfl98} preclude his unambiguously
distinguishing more states than the dimension of this space. Even to
distinguish $\bar DD$ cases unambiguously, each with some positive probability
of success, he is forced to use a POVM of the form
\begin{equation}
  B_y = |B_y\rgl\lgl B_y| \text{ for } 1\leq y\leq \bar DD,\quad B_0 = I_c\ot
I_b - \sum_{y\geq 1} B_y,
\label{eqn32}
\end{equation}
where each $|B_y\rgl$ is an element of $\HS_c\ot\HS_b$.

For her part, Alice must be able to prepare for each $x$ a \emph{pure} state
$|C_x\rgl\in\HS_c\ot\HS_b$, and these states, which in general will not be
orthogonal for different $x$, must form a basis of the space.  In the language
of the general encoding protocol, Sec.~\ref{sct2}, production of pure states
means that for each $x$ the Kraus rank of the corresponding operation is 1, so
one only needs a single term in the sum in \eqref{eqn4}.  See App.~\ref{sctpa}
for the proof of this intuitively obvious result. As a consequence, the
normalization condition \eqref{eqn5} becomes
\begin{equation}
  A_x\ad A_x=I_a,
\label{eqn33}
\end{equation}
where, as in Sec.~\ref{sct2}, we omit the redundant $l$ in the
subscript.  This means that $A_x$ is unitary for $\bar D=D$ and an isometry for
$\bar D < D$.  In short, unambiguous dense coding of $\bar DD$ messages (with,
of course, $\bar D \leq D$) means the encoding \emph{must} be isometric
(unitary for $\bar D=D$); other possibilities are excluded.

Finally,
\eqref{eqn6} translates into the condition
\begin{equation}
  \lgl B_y|C_x\rgl = \lgl B_y|\lp A_x\ot I_b\rp |\Phi\rgl
=\sqrt{\tau_x}\,\dl_{xy}.
\label{eqn34}
\end{equation}
Here $|C_x\rgl$ is normalized, since $|\Phi\rgl$ is normalized and $A_x\ot I_b$
is an isometry, whereas the $|B_y\rgl$ states are not normalized, but satisfy
the inequality
\begin{equation}
  \sum_{y\geq 1} |B_y\rgl\lgl B_y| \leq I_c\ot I_b,
\label{eqn35}
\end{equation}
which is necessary so that $B_0\geq 0$ in \eqref{eqn32}. (One can always choose
the phases so that the inner products in \eqref{eqn34} are positive.)

At this point we find it convenient to reformulate the problem slightly using
map-state duality (see \cite{ZcBn04,Grff05,GWYC05}). Let us ``transpose'' (as
that term is used in \cite{GWYC05}) $A_x$ in the form \eqref{eqn17} into the
ket
\begin{equation}
  |A_x\rgl = \sum_j |\gm^j_x\rgl\ot |a_j\rgl\in \HS_c\ot\HS_a,
\label{eqn36}
\end{equation}
and $|\Phi\rgl$ in \eqref{eqn1} into the nonsingular map $\hat\Phi:
\HS_a\ra\HS_b$ defined by 
\begin{equation}
  \hat \Phi = \sum_j \lm_j |b^j\rgl\lgl a^j|.
\label{eqn37}
\end{equation}
These transpositions allow one to rewrite \eqref{eqn34} in the equivalent form
\begin{equation}
  \lgl B_y|\blp I_c\ot\hat \Phi\brp|A_x\rgl = \sqrt{\tau_x}\, \dl_{yx}.
\label{eqn38}
\end{equation}

A little thought will show that \eqref{eqn38} can only be satisfied for all $x$
and $y$ between 1 and $\bar DD$, with $\tau_x>0$ for every $x$, if the
$\{|A_x\rgl\}$ are linearly independent and hence form a basis of
$\HS_c\ot\HS_a$, and likewise the $\{|B_y\rgl\}$ form a basis of
$\HS_c\ot\HS_b$.  Because $\{|B_y\rgl\}$ is a basis, it has a unique dual or
reciprocal basis (e.g., Sec.~15 of \cite{Hlms58}) $\{|\bar B_y\rgl\}$
satisfying
\begin{equation}
  \lgl\bar B_y|B_x\rgl = \lgl B_x|\bar B_y\rgl = \dl_{yx}.
\label{eqn39}
\end{equation}
Multiplying \eqref{eqn38} on both sides by $|\bar B_y\rgl$, summing over $y$,
and using the fact that
\begin{equation}
  \sum_y |\bar B_y\rgl\lgl B_y| = I_b\ot I_c,
\label{eqn40}
\end{equation}
yields the expression
\begin{equation}
  \blp I_c\ot\hat \Phi\brp |A_x\rgl = \sqrt{\tau_x}\,|\bar B_x\rgl.
\label{eqn41}
\end{equation}
connecting Alice's operations to Bob's measurements.  Similarly, with $\{|\bar
A_x\rgl\}$ the dual basis to $\{|A_x\rgl\}$,
\begin{equation}
  \blp I_c\ot\hat \Phi\ad\brp |B_x\rgl = \sqrt{\tau_x}\,|\bar A_x\rgl.
\label{eqn42}
\end{equation}

Since $\hat\Phi$ has an inverse, one can rewrite \eqref{eqn41} in the form
\begin{equation}
  \tau_x^{-1/2} |A_x\rgl = \blp I_c\ot \hat \Phi^{-1} \brp |\bar B_x\rgl,
\label{eqn43}
\end{equation}
and from this and from
\begin{equation}
  \sum_{y\geq 1} |\bar B_y\rgl\lgl \bar B_y| \geq I_c\ot I_b,
\label{eqn44}
\end{equation}
which is equivalent to \eqref{eqn35}, obtain the inequality
\begin{equation}
  \sum_x (1/\tau_x)\cdot |A_x\rgl\lgl A_x| \geq I_c\ot
\blp\hat\Phi\ad\hat\Phi\brp^{-1}.
\label{eqn45}
\end{equation}
Tracing both sides over $\HS_c$ and using the fact that
\begin{equation}
  \Tr_c\blp|A_x\rgl\lgl A_x|\brp = A_x\ad A_x = I_a,
\label{eqn46}
\end{equation}
one arrives at
\begin{equation}
\Blp\sum_x 1/\tau_x\Brp I_a \geq
  D\blp\hat \Phi\ad \hat \Phi\brp^{-1},
\label{eqn47}
\end{equation}
which is equivalent to
\begin{equation}
  \lgl 1/\tau\rgl := (1/\bar DD)\sum_x (1/\tau_x) \geq (\lm_{\bar D}^2\bar
D)^{-1},
\label{eqn48}
\end{equation}
since $\lm_{\bar D}$ is the smallest Schmidt coefficient of $|\Phi\rgl$. In
particular, if $\tau_x=P_c$ is a \emph{constant} independent of $x$,
\eqref{eqn48} tells us that the success probability $P_s=P_c$ is bounded by
\begin{equation}
  P_c \leq \bar D\lm_{\bar D}^2.
\label{eqn49}
\end{equation}

	\section{Orthogonal Isometries}
\label{sct6}

	\subsection{Deriving a bound for $P_s$}
\label{sct6a}

As shown in Sec.~\ref{sct5}, if for $\bar D\leq D$ a full set of $\bar DD$
messages are to be sent unambiguously with positive probabilities, Alice's
encoding operation must be an isometry. We now make the much stronger
assumption that the collection $\{A_x\}$ of isometries used for encoding is
orthogonal in the sense of \eqref{eqn16}, which is equivalent to
\begin{equation}
  \lgl A_x|A_y\rgl = \bar D\dl_{xy}
\label{eqn50}
\end{equation}
for the corresponding kets defined in \eqref{eqn36}.  This means that the
elements of the dual basis are given by
\begin{equation}
  |\bar A_x\rgl = (1/\bar D)|A_x\rgl,
\label{eqn51}
\end{equation}
and combining this with \eqref{eqn42} yields the expression
\begin{equation}
  \blp I_c\ot\hat\Phi\ad\brp\blp |B_x\rgl\lgl B_x|\brp\blp I_c\ot\hat\Phi\brp =
\blp\tau_x/\bar D^2\brp |A_x\rgl\lgl A_x|.
\label{eqn52}
\end{equation}
Sum both sides over $x$ and use the inequality \eqref{eqn35} to obtain
\begin{equation}
  \sum_x\blp\tau_x/\bar D^2\brp |A_x\rgl\lgl A_x|\leq I_c\ot \hat \Phi\ad \hat
\Phi.
\label{eqn53}
\end{equation}

Tracing both sides over $\HS_c$ and using \eqref{eqn46} yields the inequality
\begin{equation}
  \Blp\sum_x\tau_x\Brp I_a\leq \bar D^2 D \hat \Phi\ad \hat \Phi.
\label{eqn54}
\end{equation}
Because the smallest eigenvalue of $\hat \Phi\ad \hat \Phi$ is $\lm_{\bar
D}^2$, this means that
\begin{equation}
  P_s = \lgl\tau\rgl = (1/\bar DD)\sum_x\tau_x \leq \lm_{\bar D}^2\bar D,
\label{eqn55}
\end{equation}
which can be compared with \eqref{eqn48}. Note that while
\eqref{eqn48} holds quite generally for saturated unambiguous dense coding,
the derivation of \eqref{eqn55} requires the additional orthogonality
assumption \eqref{eqn16} or \eqref{eqn50}.  In the particular case in which
$\tau_x=P_c$ is a constant independent of $x$, both \eqref{eqn48} and
\eqref{eqn55} lead to the same bound \eqref{eqn49}.

The bound \eqref{eqn55} is tighter for the case $L=\bar DD$ than the
information-theoretic bound \eqref{eqn14}, as can be seen by writing the latter
in the form
\begin{equation}
  P_s \leq \frac{\log D + H_E\mspace{15mu}}{\log D + \log\bar D}.
\label{eqn56}
\end{equation}
As noted in part B below, the entangled state $|\Phi\rgl$ can with a
probability $\lm_{\bar D}^2\bar D$ be transformed into a uniformly entangled
state with entanglement $\log\bar D$ by a local operation, and since such an
operation cannot increase the average entanglement \cite{Bnao96}, it follows
that
\begin{equation}
  \lm_{\bar D}^2\bar D\log\bar D \leq H_E.
\label{eqn57}
\end{equation}
Using this and the fact that $H_E$ cannot exceed $\log\bar D$ it is
straightforward to show that the right side of \eqref{eqn55} is bounded above
by the right side of \eqref{eqn56}.

	\subsection{Saturating the bound}
\label{sct6b}

In fact, given \emph{any} collection of $\bar DD$ orthogonal isometries, there
is a POVM of the form \eqref{eqn32} which results in each message being
transmitted with the same probability of success $\tau_x=\lm_{\bar D}^2\bar D$,
saturating the bound \eqref{eqn55}.  The easy way to see this is to imagine Bob
carrying out his part of the protocol in two steps: an unambiguous entanglement
concentration operation (the term used in \cite{GWYC05} for what its
originators \cite{BBPS96} called the ``Procrustean method'') on his particle,
which if it succeeds transforms $|\Phi\rgl$ into a uniformly entangled state
with the same Schmidt rank, followed by a measurement, in an orthonormal basis
of the type described in Sec.~\ref{sct4a}, on the combined system of
his particle and the one received from Alice.

Unambiguous entanglement concentration results from Bob carrying out an
operation \cite{iKrs83,iNlCh00} described by Kraus operators
\begin{equation}
  K_1=\sum_j\blp\lm_{\bar D}/\lm_j\brp |b^j\rgl\lgl b^j|,\quad K_2 =
\sqrt{I_b-K_1\ad K_1}\,.
\label{eqn58}
\end{equation}
It is successful if $K_1$ occurs, for which the probability is
\begin{equation}
  \lgl\Phi| K_1\ad K_1 |\Phi\rgl = \lm_{\bar D}^2\bar D,
\label{eqn59}
\end{equation}
and then Alice and Bob share the uniformly entangled state $(1/\sqrt{\bar
D})\sum_j |a^j\rgl\ot|b^j\rgl$.

The dense coding protocol consists of Alice encoding in the manner indicated in
Sec.~\ref{sct4a}, and Bob attempting unambiguous entanglement concentration in
the manner just described.  If the latter is successful, Bob carries out
projective measurements on the two particles as described in Sec.~\ref{sct4a},
certain that the outcome accurately reflects Alice's encoding.  Obviously, the
probability of success \eqref{eqn59} is independent of $x$.  The two steps of
entanglement concentration followed by projective measurement can be combined
into a single POVM of the form \eqref{eqn32} by setting $|B_x\rgl =
K_1|\Phi_x\rgl$, with $|\Phi_x\rgl$ defined in \eqref{eqn19}.  It is then a
simple exercise to show that the inequality \eqref{eqn35} is satisfied.

	\subsection{Exceeding the bound}
\label{sct6c}

The inequality \eqref{eqn55} was derived by requiring that all $\bar DD$
messages be transmitted with positive probability and that the isometries used
for encoding be orthogonal, so it is interesting to ask whether it holds if
either condition is relaxed.  We believe that both are necessary, but have not
been able to prove this.  Some insight is, however, provided by the following
considerations.  Suppose that $\bar D$ is 3 or more, and the $\lm_j$ in
\eqref{eqn1} are all equal, except for $\lm_{\bar D}=\ep>0$, which is much
smaller than the others.  Then Bob can carry out an operation analogous to
\eqref{eqn58}, but with 
\begin{equation}
  K_1 = \sum_{j=1}^{\bar D-1} |b^j\rgl\lgl b^j|,\quad K_2 = |b^{\bar D}\rgl\lgl
b^{\bar D}|,
\label{eqn60}
\end{equation}
which one can think of as a projective measurement to determine whether or not
his particle is in $|b^{\bar D}\rgl$. If, with probability $1-\ep^2$, $K_1$
occurs, the resulting uniformly-entangled state can be used to transmit $D(\bar
D-1)$ messages in a deterministic manner, using the protocol in
Sec.~\ref{sct4a}.  Then $\lgl\tau\rgl$ as defined in \eqref{eqn55}, assuming
$\tau_x=0$ for $x>D(\bar D-1)$, is $(1-\ep^2)(1-1/\bar D)$, which can obviously
be made larger than $\ep^2\bar D$, which is the right side of \eqref{eqn55}.

This example violates both of the conditions used to derive \eqref{eqn55},
since we no longer have saturation---$\tau_x=0$ for some of the messages---and
the isometries used for encoding now map a $(\bar D-1)$-dimensional Hilbert
space onto one of $D$ dimensions.  While they can be extended to isometries
acting on the original $\bar D$-dimensional space $\HS_a$, these isometries
will not be orthogonal, see App.~\ref{sctpb}.  Thus the possibility
remains open that \eqref{eqn55} might be valid given the assumption of one or
the other but not both of the conditions used to derive it, though we ourselves
doubt that this is the case.  By contrast, the inequality
\eqref{eqn48} is known to hold for a saturated protocol, but is obviously
useless if some of the $\tau_x$ are zero.

	\section{Unambiguous Dense Coding for $\bar D>D$}
\label{sct7}

In contrast to the situation discussed in Secs.~\ref{sct5} and \ref{sct6}, in
which the Schmidt rank of the entangled state is less than or equal to the
dimension of the quantum channel, $\bar D \leq D$, we have very few results for
unambiguous dense coding if $\bar D$ is greater than $D$; in particular, we
have no upper bounds on success probabilities analogous to those in
\eqref{eqn48}, \eqref{eqn49} and \eqref{eqn55}.  The situation is not unlike
that for deterministic dense coding with $\bar D > D$ as discussed in
Sec.~\ref{sct4c}: we have more questions than answers.

There is a simple argument that shows that with $\bar D > D$ it is
always possible to send $D^2$ messages in an unambiguous fashion, and one
can place lower limits on the probability of success. Assume, as
previously, that the Schmidt coefficients are arranged in decreasing order,
\eqref{eqn2}, and define the two projectors 
\begin{equation}
  Q_a = \sum_{j=1}^D |a^j\rgl\lgl a^j|,\quad 
  Q_b =  \sum_{j=1}^D |b^j\rgl\lgl b^j|
\label{eqn61}
\end{equation}
on $\HS_a$ and $\HS_b$.  Then $\{Q_a,\tilde Q_a=I_a-Q_a\}$ and $\{Q_b,\tilde
Q_b=I_b-Q_b\}$ form projective decompositions of the identities $I_a$ and
$I_b$.  If Alice and Bob carry out projective measurements using these
decompositions, it is evident from \eqref{eqn1} that their results will be
perfectly correlated, and outcome $Q_a$ will be accompanied by $Q_b$ with
probability
\begin{equation}
  P_m = \sum_{j=1}^D (\lm_j)^2.
\label{eqn62}
\end{equation}  
When this occurs, Alice and Bob share a partially entangled state of Schmidt
rank $D$, of the form \eqref{eqn1}, but with the summation limit replaced by
$D$, and $\lm_j$ replaced by $\hat\lm_j=\lm_j/\sqrt{P_m}$. This entangled state
can be used for unambiguous transmission of $D^2$ messages and our discussion
in Secs.~\ref{sct5} and \ref{sct6} applies, provided the success probabilities
calculated using the $\{\hat\lm_j\}$ are at the end multiplied by $P_m$.  In
particular we have an overall protocol, by combining the projective measurement
just discussed with the entanglement concentration of Sec.~\ref{sct6b}, in
which allows each of $D^2$ messages to be sent with a probability of success
equal to $D\lm_D^2 P_m$.  This might be optimal for the equal-probability case,
but we have no proof that it is.

Alice's projective measurement is actually not necessary in this scheme, since
she can always proceed as if the measurement would have been successful, and
leave it to Bob to declare the transmission a failure if his outcome is $\tilde
Q_b$.  On the other hand it will not do for Alice alone to carry out the
measurement and communicate the result to Bob, since this requires a
classical side channel (or something similar), and lies outside the scope of
protocols we are considering.

Having both parties carry out the projective measurement has the additional
advantage that if the common outcome corresponds to $\tilde Q_a$ and $\tilde
Q_b$, there will still be some entanglement left, if $\bar D - D\geq 2$, or at
the very least the bare quantum channel when $\bar D=D+1$, which can be used to
communicate some messages. These considerations, while of some interest, tell
us very little about possible optimal protocols.  We do not even know if $D^2$
is an upper bound on the number of messages that can be sent unambiguously when
$\bar D$ is greater than $D$, though we suspect this is the case.

	\section{Comments on Previous Work}
\label{sct8}

	\subsection{Uniformly entangled states for $\bar D<D$}
\label{sct8a}

The system of orthogonal isometries for $\bar D<D$ in Sec.~\ref{sct4a} was
first proposed, so far as we are aware, in Sec.~3 of \cite{BrEk95}, and later
worked out independently in \cite{YnWn04}.  In both papers it is assumed that
Alice's and Bob's particles have Hilbert spaces of different dimensions, $D$
and $\bar D$ in our notation, and are in a uniformly entangled state of Schmidt
rank equal to the smaller dimension.  The encoding operation is thought of as a
unitary carried out on the space of higher dimension.  While there is nothing
wrong with discussing encoding using unitaries rather than isometries---see the
comments following \eqref{eqn5} in Sec.~\ref{sct2}---it can give rise to
confusion, because while the encoding task dictates the nature of the
isometries, their extension to unitaries acting on a range space of higher
dimension is to a large measure arbitrary.  Such confusion may lie behind the
incorrect definition of the unitary operator $L_t$ and the incorrect trace
formula (12) in \cite{BrEk95}, both of which are valid (aside from a
typographical error) for $\bar D=D$, but not for $\bar D < D$, and an
incorrect expression (11) for the unitary operator $U_{mn}$ in
\cite{YnWn04}.

	\subsection{Deterministic dense coding}
\label{sct8b}

A very interesting exploration of deterministic dense coding for small systems
of partially-entangled pure states has been carried out using a combination of
numerical and analytic techniques by Mozes et al. \cite{MzOR05}; some of their
results extend to general $D$ (in our notation), and they make interesting
conjectures about the $D$-dependence of others.  They assume without discussing
the matter that unitary encoding is optimal, but do not assume that the
unitaries are orthogonal in the sense in which we use that term, see
\eqref{eqn16}.  (The term ``orthogonal'' in their paper has a different
meaning.)

It is of particular interest that our inequality \eqref{eqn26}, which
transcribed to their notation reads $\lm_0 N_\text{max} \leq d$, is saturated
in a number of cases by their numerical or analytical results or conjectures,
in the sense that there are nontrivial choices of entangled states for which
this inequality is an equality.  These include the right-most limits of the
regions 5, 6, and 7 in their Fig.~1---the numbers refer to $L_d$
($N_\text{max}$)---for $D=3$, and their conjectured minimal-entanglement states
for $L_d=D+n$ for $n=2,\,3,\ldots D$.  It is interesting that the $L_d=7$ case
for $D=3$ is not included in their general conjectures.

Another point of agreement between their work and ours is their observation
that $L_d$ is not in general a monotone increasing function of the
entanglement. Our bound on $L_d$ depends on $\lm_1$, not on the entanglement.

	\subsection{Unambiguous dense coding}
\label{sct8c}

The earliest work on unambiguous dense coding known to us is the study of Hao
et al.\ \cite{HaLG00} of a partially entangled state of two qubits.  (This and
and the later \cite{PtPA05} use the term ``probabilistic dense coding.'')  The
paper actually includes two schemes. The first requires a classical side
channel, but the second does not, and thus fits within the framework of our
discussion.  The encoding scheme employs orthogonal unitaries, and the
probability of success saturates the bound \eqref{eqn55}, in agreement with our
Sec.~\ref{sct6b}.

More recently Pati et al.\ \cite{PtPA05} have independently worked out the
qubit case, with results in agreement with \cite{HaLG00}. They also considered
its extension to general $D$, using a particular collection of $D^2$ orthogonal
unitaries, assuming Alice and Bob share a partially entangled pure state.
Their upper bound for the success probability has now been superseded by our
\eqref{eqn55}; the latter is both a tighter bound (in some sense the best
possible---see Sec.~\ref{sct6b}), and was derived under weaker assumptions.

In addition, these authors construct an example in which an increase in
entanglement brought about by increasing the Hilbert space dimensions of both
Alice's and Bob's particles can give rise to a lower average probability for
transmitting a given collection of messages, even when a uniformly entangled
state is employed.  We believe it is best to think of this somewhat
counterintuitive result as arising from the encoding scheme they propose for
the larger system; in particular, its unitaries are no longer orthogonal.
While there is no reason to suppose that greater entanglement will always
improve a dense coding scheme---see the comments in \cite{MzOR05} regarding the
deterministic case---we think the main lesson to be drawn from the example
considered in \cite{PtPA05} is the importance of paying attention to the
encoding process, not just optimizing Bob's measurements.

	\subsection{Unambiguous state discrimination}
\label{sct8d}

Unambiguous dense coding is related to unambiguous state discrimination,
see \cite{Chfl98,ChBr98,SnHB01}, in the sense that Bob's measurement task is to
distinguish the two-particle states $\{|C_x\rgl\}$, see \eqref{eqn34}, in an
optimal fashion.  In the case of dense coding these states are somewhat special
in that they all correspond to the same reduced density operator on the Hilbert
space $\HS_b$ of Bob's particle.  In addition, whereas in unambiguous state
discrimination one is generally concerned with optimal discrimination of a set
of states thought of as simply given in advance, in the dense coding case
optimization involves Alice's choice of operations for producing the
$\{|C_x\rgl\}$ as well as Bob's choice of a POVM to distinguish them.

Some of our results are related to previous work on unambiguous state
discrimination in the following way.  When the success probabilities of state
discrimination are required to be equal, Chefles in \cite{Chfl98} derived an
optimal average success probability consistent with our \eqref{eqn49}.  Also if
in Sec.~\ref{sct6} when $\bar D=D$ the orthogonal unitary operators are of the
special form used in \cite{PtPA05}, then the states to be distinguished are
divided into $D$ mutually orthogonal sets, and the states in each set are
linearly independent and symmetric. Therefore, the solution given in
\cite{ChBr98} for the optimal discrimination of symmetric states can be used to
obtain the maximum average success probability in
\eqref{eqn55}. However, our result is more general in the sense that it does
not depend on any specific form of orthogonal unitaries.

	\subsection{Noisy entangled states}
\label{sct8e}

While dense coding using noisy (mixed) entangled states lies outside the scope
of the research reported in this paper, there is one feature of the
studies of this problem in 
\cite{BsPV00,BsVd00,Bwn01,Bwn01b,Hrsh01,Bn02,Zmzk03,Brao04} to which we wish
to draw attention.  These papers arrive at a rather simple formula
\begin{equation}
  C = \log D + S(\rho_B) -S(\rho)
\label{eqn63}
\end{equation}
for the optimal asymptotic classical capacity, with $D$ the dimension of the
noiseless quantum channel, $\rho$ the density operator of the initial entangled
state, $\rho_b$ its partial trace down to Bob's particle, and $S$ the von
Neumann entropy.  This result is derived assuming that Alice is restricted to
\emph{unitary} encoding of messages, whereas Bob is allowed, and in general
must employ, the most general decoding operation, including coherent
measurements on states resulting from multiple transmissions.

There is no reason why $S(\rho)$ cannot be larger than $S(\rho_b)$---an extreme
example is a maximally-mixed state---and if that is the case, \eqref{eqn63}
\emph{cannot} be the optimal capacity, since $C=\log D$ is always possible by
throwing away the entangled state and using the quantum channel in a
straightforward way to transmit $D$ messages.  To be sure, it is conceivable
that \eqref{eqn63} might hold whenever $S(\rho_b)$ exceeds $S(\rho)$, i.e.,
when the right side is greater than $\log D$, but we know of no compelling or
even plausible argument to this effect.  What is clearly needed is a study of
what can be achieved using alternative methods of encoding, and until that has
been carried out it seems best to regard \eqref{eqn63} as a lower bound for,
rather than the actual value of, the optimal capacity for a mixed entangled
state. See the additional comments in Sec.~\ref{sct9b}.

	\section{Conclusion}
\label{sct9}

	\subsection{Summary}
\label{sct9a}

We studied the problem of dense coding using a partially-entangled pure
state whose Schmidt rank $\bar D$ can be different from the dimension $D$ of
the noiseless quantum channel used to communicate from Alice to Bob, both for
deterministic protocols in which a maximum of $L_d$ messages can be sent with
perfect fidelity, and also for unambiguous protocols in which message $x$
is faithfully transmitted with a probability $\tau_x$. 

In the deterministic case we considered uniformly-entangled states for $\bar D<
D$, where for completeness the previously published encoding protocol for
sending $L_d=\bar DD$ messages was included in Sec.~\ref{sct4a}, and for $\bar
D > D$, where we showed in Sec.~\ref{sct4c} that $L_d$ is actually less than
$D^2$, unless $\bar D$ is a multiple of $D$, and if it is a multiple of $D$
other states besides one that is uniformly entangled can be used to achieve the
optimal protocol. For pure states that are not uniformly entangled, our
principal result is the inequality \eqref{eqn26} bounding $L_d$ in terms of the
largest Schmidt coefficient $\lm_1$.  The utility of this bound is confirmed by
the fact that it is satisfied as an equality by several results and conjectures
in \cite{MzOR05}, as discussed in Sec.~\ref{sct8b}.  We also showed in
Sec.~\ref{sct4d} that a protocol which achieves $L_d$ cannot be used to send
additional messages in an unambiguous fashion, i.e., with a nonzero probability
of failure.

For unambiguous protocols, our main results are for $\bar D\leq D$ and the
saturated case in which $\tau_x>0$ for all $\bar D D$ messages.  This implies
that the encoding operation must be an isometry, and that allows an analysis in
Sec.~\ref{sct5} leading to the inequality \eqref{eqn48} which bounds the
average value of $1/\tau_x$ in terms of the smallest Schmidt coefficient
$\lm_{\bar D}$.  If in addition one assumes the isometries are mutually
orthogonal, there is an analogous bound \eqref{eqn55} on the average of
$\tau_x$, that is, the average probability of success.  For the case in which
all the $\tau_x$ are identical both of these inequalities yield the same bound,
and we showed that it can be achieved by a protocol in which Bob first uses
entanglement concentration to produce, with some probability, a uniformly
entangled state. If successful this allows use of the dense coding protocol for
such states, as described in Sec.~\ref{sct4a}.  We also showed that for 
$\bar D > D$ it is always possible to send $D^2$ messages unambiguously, though
we have no bounds comparable to those for $\bar D\leq D$.

	\subsection{Open questions}
\label{sct9b}

In the case of deterministic dense coding a very significant and challenging
problem is to determine the ``phase diagram'' of the maximum number $L_d$ of
messages as a function of the Schmidt coefficients $\{\lm_j\}$ of the entangled
state.  The work in \cite{MzOR05} represents a good beginning, and it would be
of interest to confirm the conjectures given there, and extend them if possible
to more general principles determining, or at least strongly constraining, the
values of $L_d$.  Since our inequality \eqref{eqn26} seems satisfied as an
equality at certain boundary points of their $L_d$ regions, one can hope that
it, or perhaps other inequalities yet to be discovered, might prove of some
assistance in working out the phase diagram.  A major unanswered question is
whether isometric (unitary) encoding is always sufficient to achieve $L_d$, or
whether for certain entangled states one needs to employ a more general
encoding procedure.  Could it be, for example, that the absence of cases in
which $L_d=D^2-1$ reported in \cite{MzOR05} reflects a limitation of unitary
encoding?  Only further study can answer that and similar questions.  The case
of $\bar D> D$, where isometric encoding is clearly not possible, is even less
well understood than that of $\bar D\leq D$, apart from certain special
entangled states in the case where $\bar D$ is an integer multiple of $D$.
Even for the simple example of $\bar D=3$ and $D=2$, we do not know whether it
is possible by dense coding to send more than the $D=2$ deterministic messages
allowed by the quantum channel itself.

While we have identified an optimal protocol for unambiguous dense coding when
the success probability $\tau_x$ is independent of $x$ for all $\bar D D$
messages, assuming $\bar D\leq D$, the case in which $\tau_x$ depends on $x$
remains open; we do not even have a bound on the average probability of
success, aside from the information-theoretic \eqref{eqn14}.  The bound
\eqref{eqn48} on $\lgl 1/\tau\rgl$ requires $\tau_x>0$ for all $x$, and is
unlikely to be helpful when some of these probabilities are quite small.  If
some of the $\tau_x$ are zero one needs to allow for the possibility of
nonunitary or nonisometric encoding, which may be a difficult task.  We know
even less about what can be done unambiguously when $\bar D$ is larger than
$D$.  It is always possible to send $D^2$ messages unambiguously with positive
probabilities, but is this the largest number possible?  We suspect it is, but
it would be nice to have a proof. 

Both in the deterministic and in the unambiguous case, and also for noisy
entangled states as discussed in Sec.~\ref{sct8d}, a major unanswered question
is whether optimal protocols can always be based on unitary encoding, or
whether more general procedures are sometimes needed.  Up till now almost all
studies of extensions of standard dense coding have simply assumed unitary (or
isometric) operations.  But in very few cases are there proofs or even
plausible arguments that this type of encoding is optimal.  Of course, when
studying a difficult problem it is often a good strategy to make various
assumptions which allow one to calculate an explicit result, and leave till
later the problem of justifying them. We in no way wish to undervalue work of
this type (including our own) and the insights which have emerged from it.
Nonetheless, one should not forget that more general forms of encoding exist,
and that in particular situations they might lead to better transmission of
information.  Exploring the encoding process is itself an interesting problem
which might make a significant addition to our understanding of the basic
quantum principles underlying dense coding.

	\section*{Acknowledgments} We thank S. Mozes for helpful correspondence
concerning \cite{MzOR05}, and an anonymous referee for encouraging us to add
the material in Sec.~\ref{sct7}.  The research described here received support
from the National Science Foundation through Grants PHY-0139974 and
PHY-0456951, and one of us (Wu) acknowledges support from the Chinese Academy
of Sciences and the University of Science and Technology of China.

\appendix
\numberwithin{equation}{section}

	\section{Appendix. Kraus Rank and Density Operator}
\label{sctpa}

We will show that for the framework used in Sec.~\ref{sct2}, see
Fig.~\ref{fgr1}, the Kraus rank of the encoding operation \eqref{eqn4}, which
is to say the number of linearly independent $\{A_{xl}\}$ operators for a fixed
$x$, is equal to the rank of the density operator $\rho_{cb}$ describing the
state available to Bob for measurements.  In what follows, $x$ is always fixed.
It could be omitted from the notation, but retaining it clarifies the
connection with formulas in Sec.~\ref{sct2}.
  
Let the state resulting from unitary time evolution, Fig.~\ref{fgr1}, be
\begin{equation}
  |\Psi\rgl = \blp W_x\ot I_b\brp\blp |g^0\rgl\ot|\Phi\rgl\brp
=\sum_{jl}\lm_j|h^l\rgl\ot\blp A_{xl}|a^j\rgl\brp\ot|b^j\rgl.
\label{aeqn1}
\end{equation}
It gives rise to a reduced density operator
\begin{equation}
  \rho_{cb} = \Tr_h\blp |\Psi\rgl\lgl\Psi|\brp = MM\ad,
\label{aeqn2}
\end{equation}
where
\begin{equation}
  M =\sum_{jl}\lm_j\left[ A_{xl}|a^j\rgl\ot|b^j\rgl\right]\lgl h^l| =\blp
I_c\ot \hat \Phi\brp K
\label{aeqn3}
\end{equation}
is a map from $\HS_h$ to $\HS_c\ot\HS_b$, $\hat\Phi$ is defined in
\eqref{eqn37}, and
\begin{equation}
  K=\sum_{jl}\left[ A_{xl}|a^j\rgl\ot|a^j\rgl\right]\lgl h^l|,
\label{aeqn4}
\end{equation}
maps $\HS_h$ to $\HS_c\ot\HS_a$.
The argument that identifies the Kraus rank with the rank of $\rho_{cb}$
proceeds in three steps: (i) the rank of $K$ is the Kraus rank; (ii) the rank
of $M$ is that of $K$; (iii) the rank of $MM\ad$ is the rank of $M$.

For the first step, note that if $\{|c^m\rgl\}$ is an orthonormal basis of
$\HS_c$, the matrix elements of $K$ are
\begin{equation}
  \lgl c^m a^j|K|h^l\rgl = \lgl c^m|A_{xl}|a^j\rgl.
\label{aeqn5}
\end{equation}
By definition, the rank of $K$ is the number of linearly independent column
vectors, labeled by $l$, in its matrix, but this is obviously the same as the
number of linearly independent matrices $\lgl c^m|A_{xl}|a^j\rgl$, again
labeled by $l$, which by definition is the Kraus rank.

That the ranks of $M$ and $K$ are the same is a consequence of the fact that
$I_c\ot\hat \Phi$ in \eqref{aeqn3} is a nonsingular map (rank $\bar DD$) from
$\HS_c\ot\HS_a$ to $\HS_c\ot\HS_b$.  That multiplying a matrix by a nonsingular
matrix leaves its rank unchanged is a standard result of linear algebra, as is
the equality of the ranks of $M$ and $MM\ad$; see, for example, p.~13 of
\cite{HrJh85}.  This completes the argument.

What is perhaps a more intuitive approach to this result can be given using
atemporal diagrams; see the very similar argument in Sec.~VC of \cite{GWYC05}
with reference to Fig.~13 of that paper.  Translated to the present context,
the essential observation is that $W_x$ in Fig.~\ref{fgr1} with $|g^0\rgl$
fixed may be considered a map from $\HS_h$ to $\HS_c\ot\HS_a$---thus $K$ as
defined in \eqref{aeqn4}---and the Kraus rank can be identified with the rank
of this ``cross operator.''

	\section{Appendix. Extensions of Orthogonal Isometries}
\label{sctpb}

Lemma: Suppose a set of $N$ mutually orthogonal isometries maps a
${d}$-dimensional space to a $D$-dimensional space $\HS_c$, and  ${d} < D$.
Then an extension of these to isometries that map a $({d}+1)$-dimensional space
to a $D$-dimensional space cannot preserve mutual orthogonality unless $N
\le D$.

Proof: Let the original and extended isometries be
\begin{equation}
  J_n = \sum_{k=1}^{{d}} |\Psi_k^{n}\rangle \langle k |,\quad
 K_n = \sum_{k=1}^{{d}+1} |\Psi_k^{n}\rangle \langle k |,
\label{beqn1}
\end{equation}
for $1\leq n\leq N$. Assume both sets are orthogonal in the sense of
\eqref{eqn16}, so that for $n\neq n'$
\begin{align}
  \Tr_c(J^{}_{n'} J\ad_{n}) 
  &=\sum_{k=1}^{{d}} \lgl \Psi_k^{n}|\Psi_k^{n'}\rgl=0,
\notag\\
   \Tr_c(K^{}_{n'} K\ad_{n})
  &=\sum_{k=1}^{{d}+1} \lgl \Psi_k^{n}|\Psi_k^{n'}\rgl=0,
\label{beqn2}
\end{align}
which means that
\begin{equation}
  \lgl \Psi_{{d}+1}^{n'}|\Psi_{{d}+1}^{n}\rgl=0
\label{beqn3}
\end{equation}
for unequal $n$ and $n'$ between 1 and $N$.  Since the
$\{|\Psi_{{d}}^{n}\rgl\}$ are a collection of $N$ nonzero (in fact, normalized)
states in a Hilbert space $\HS_c$ of dimension $D$, they can only be orthogonal
to each other for $N\leq D$, which proves the lemma.

Note that in the situation discussed in Sec.~\ref{sct6c}, where ${d}=\bar D-1$,
we are interested in a collection of $N=(\bar D-1)D$ orthogonal isometries, and
since $\bar D-1$ is 2 or more, the lemma shows that they cannot be extended to
orthogonal isometries mapping a $\bar D$- to a $D$-dimensional space.


\end{document}